\documentclass[aps, prb, a4paper, twocolumn, 10pt, floatfix]{revtex4-1}

\usepackage{amsfonts}
\usepackage{subfigure}
\usepackage{color}
\usepackage{lipsum}
\usepackage{float}
\usepackage{times}
\usepackage{graphicx}
\usepackage{amsfonts}
\usepackage{amssymb}
\usepackage{amsmath}
\usepackage[colorlinks=true,bookmarks=false,citecolor=blue,urlcolor=blue]{hyperref}

\def\phi{\varphi}
\def\theta{\vartheta}
\def\rho{r}

\begin{document}

\title{Judiciously distributing laser emitters to shape the desired far field patterns}

\author{Constantinos A. Valagiannopoulos and Vassilios Kovanis}
\affiliation{Department of Physics, School of Science and Technology, Nazarbayev University, 53 Qabanbay Batyr Ave, Astana, KZ-010000, Kazakhstan}

\begin{abstract}
The far-field pattern of a simple one-dimensional laser array of emitters radiating into free space is considered. In the path of investigating the inverse problem for their near fields leading to a target beam form, surprisingly we found that the result is successful when the matrix of the corresponding linear system is not well-scaled. The essence of our numerical observations is captured by an elegant inequality defining the functional range of the optical distance between two neighboring emitters. Our finding can  restrict substantially the parametric space of integrated photonic systems and simplify significantly the subsequent optimizations. 
\end{abstract}

\maketitle

\section{Introduction}
Recently, we have noticed across the photonics literature that there is robust research activity in fabrication, characterization, design and theoretical developments of large classes of optically interacting oscillators. Examples include: (a) Arrays of coherently coupled vertical cavity lasers \cite{Choquette_2017} where for the first time the potential of designing the gain/loss profile for non-Hermitian systems is experimentally demonstrated. (b) A network of 37 Quantum Cascade Lasers via antenna mutual coupling \cite{MIT_QCL_2016} which provide a path to an increased output power, while keeping the intensity of each individual laser low, with help from a suitable phase distribution. (c) A network of 64-by-64 two-dimensional (2D) large-scale nanophotonic phased array employed to produce a great variety of radiation patterns useful in several applications beyond conventional beam focusing and steering \cite{MIT_Watts_2013}. (d) Fabrication and testing of a large set of silicon optical integrated circuits and implementation of two-dimensional optical beam steering in indium phosphite-based photonic integrated circuits. \cite{UCSB_2013} (e) An array of commercially available vertical cavity lasers interacting via diffractive coupling \cite{Fisher_2015} and (f) manifestation of turbulent chimera states in phased arrays of diode lasers \cite{Ioanna}. These photonic circuit implementations are driven by technological applications, such as chip scale laser radars, short range optical network communications, next generation imaging, sensing and on-demand generation of optical diverse waveforms.

In parallel, we have also recorded substantial theoretical activity in developing an \textit{Inverse Design Paradigm Shift} in photonic design as it applies to next generation ultra-performing devices. Based on applied mathematical ideas, the so-called level set methods for computing moving fronts by Stanley Osher \cite{Osher_1988,Osher_2003}, have been recently translated to photonic devices \cite{OwenPhD} to the benefit of examining various bounds of solar cells design and other metamaterial synthetic structures \cite{Jelena_2015}. Prior to the aforementioned recent developments, notable works have also appeared in the framework of topology optimization including both configuration and materials \cite{BendsoeBook, Bendsoe}. As far as active structures are concerned, convex optimization has been employed in determining optimal currents that refined the limits for various metrics of radiation performance \cite{GustafssonCurrents, GustafssonLimit}. Similar techniques have been applied to achieve frequency-selective energy transportation with suitable mixtures of active and passive media \cite{GiantFreqSel} and for the determination of optimal dielectrics in modeling transformation-optics devices \cite{JohnsonLipsonOE}. In particular, a computationally efficient method based on dipole approximation and the reciprocity of space has been introduced by performing an iterative update for both the shape and the texture of the structure under optimization.

In this paper, motivated by such trends in coupled optical oscillators and applied mathematics methods, we inject an inverse design approach to the construction of the far field of multiple radiating optical apertures. We consider the simple case of one-dimensional array of equispaced photonic emitters and confine our study on the maximal radiation plane. We do not address how the near fields at the end of the waveguides are produced; instead, we compute the optimal near fields which, via propagation into free space, formulate the desired far-field pattern. Our key mathematical finding is that for a successful pattern reconstitution, the optical distance between two consecutive elements obeys a double inequality. The lower bound expresses the difficulty of reconstructing  a rapidly spatially varying pattern with a few number of effective sources. The upper bound is related to the essential coherence between the sources for an efficient beam forming.

This paper is organized as follows. In Section \ref{ConfAndMot}, we present the configuration of an array with several equispaced laser emitters along a line, define the output fields from the waveguides and state the assumption for two-dimensional variation. In Section \ref{Loc2Far}, the far-field pattern is deduced in terms of the local output fields of the lasers by suitably approximating the cylindrical waves emitted by the abruptly terminated waveguides in the far region. In Section \ref{DesFar2Opt}, we find the optimal set of output fields of the lasers to mimic the azimuthal variation of the target pattern and we present the range that the optical period should belong to in order to have a successful implementation of the proposed method. Finally, in the concluding Section \ref{ConcFut}, we set as future target to employ the described technique in inverse design of integrated laser systems.

\section{Configuration and Motivation}
\label{ConfAndMot}
Typically, a laser comprises a cavity of a finite-length waveguide \cite{HessPaperB}, longitudinally restricted by two mirrors which reflect the light waves and provide feedback to form an optical resonator. Into the waveguide, the gain medium gets continuously excited by the circulating electromagnetic field. However, the power into the cavity is not increasing unboundedly since one of the two mirrors is not perfect; thus, a certain portion of the signal leaks from the waveguide, which constitutes the output of the device. Very commonly, laser cavities are packed together in order to build coherently or incoherently coupled arrays \cite{HessPaperA, YarivPaper} to leverage several beneficial characteristics of the emitting beam (directivity, shaping, power etc). Such a laser array is the configuration analyzed in the work at hand.

\begin{figure}[htbp]
\centering
\includegraphics[width=9cm]{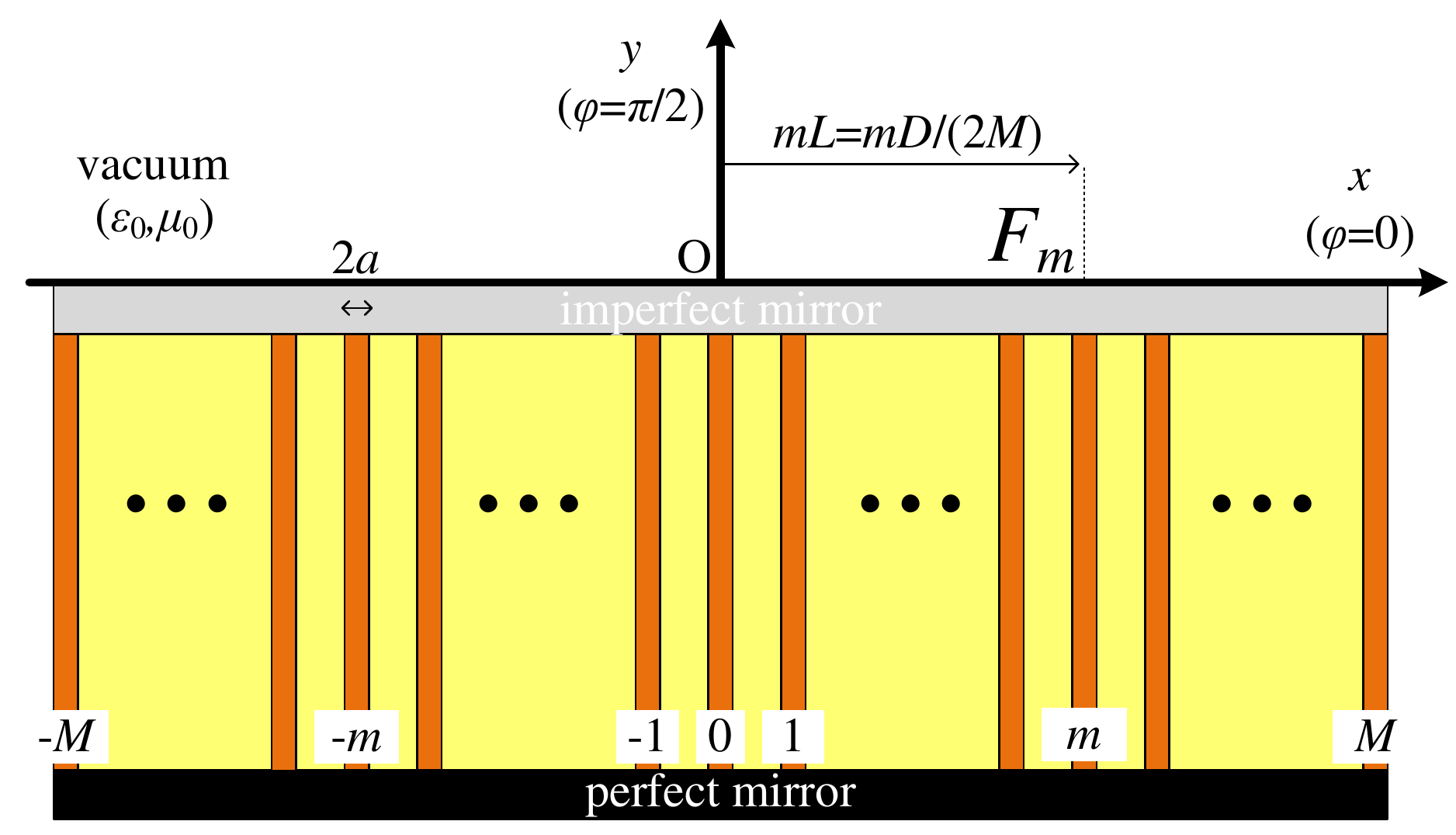}
\caption{Schematic of the configuration of the analyzed device. $(2M+1)$ laser emitters develop independently and locally at their outputs $z$-polarized electric fields $F_m$ with $m=-M,\cdots, M$, which propagate into free-space. The distance between two consecutive emitters is equal to $L=D/(2M)$, while the transversal size of each one equals 2$a$. In this way, a far-field pattern $G(\phi)$ is shaped at $y>0$.}
\label{fig:Fig1}
\end{figure}

In particular, we consider the structure schematically depicted in Fig. \ref{fig:Fig1} where $(2M+1)$ of the aforementioned laser emitters are positioned along $x$ axis and develop different $z$-polarized electric fields at their outputs, namely at the position of the imperfect mirror at $y=0$. The complex phasors of the fields are denoted by $F_m$, $m=-M,\cdots,M$, with a suppressed harmonic time $e^{+j\omega t}$. The spacing between two consecutive waveguides is given by $L$, while, with no loss of generality, the transverse size of each of them is taken equal to $2a<L$. The fields $F_m$ are considered constant throughout the cross section of each waveguide which is assumed, again with no loss of generality, circular with radius $a$ pegged at $(x,y)=(mL,0)$, $m=-M,\cdots, M$. 

\begin{figure}[htbp]
\centering
\includegraphics[width=7.3cm]{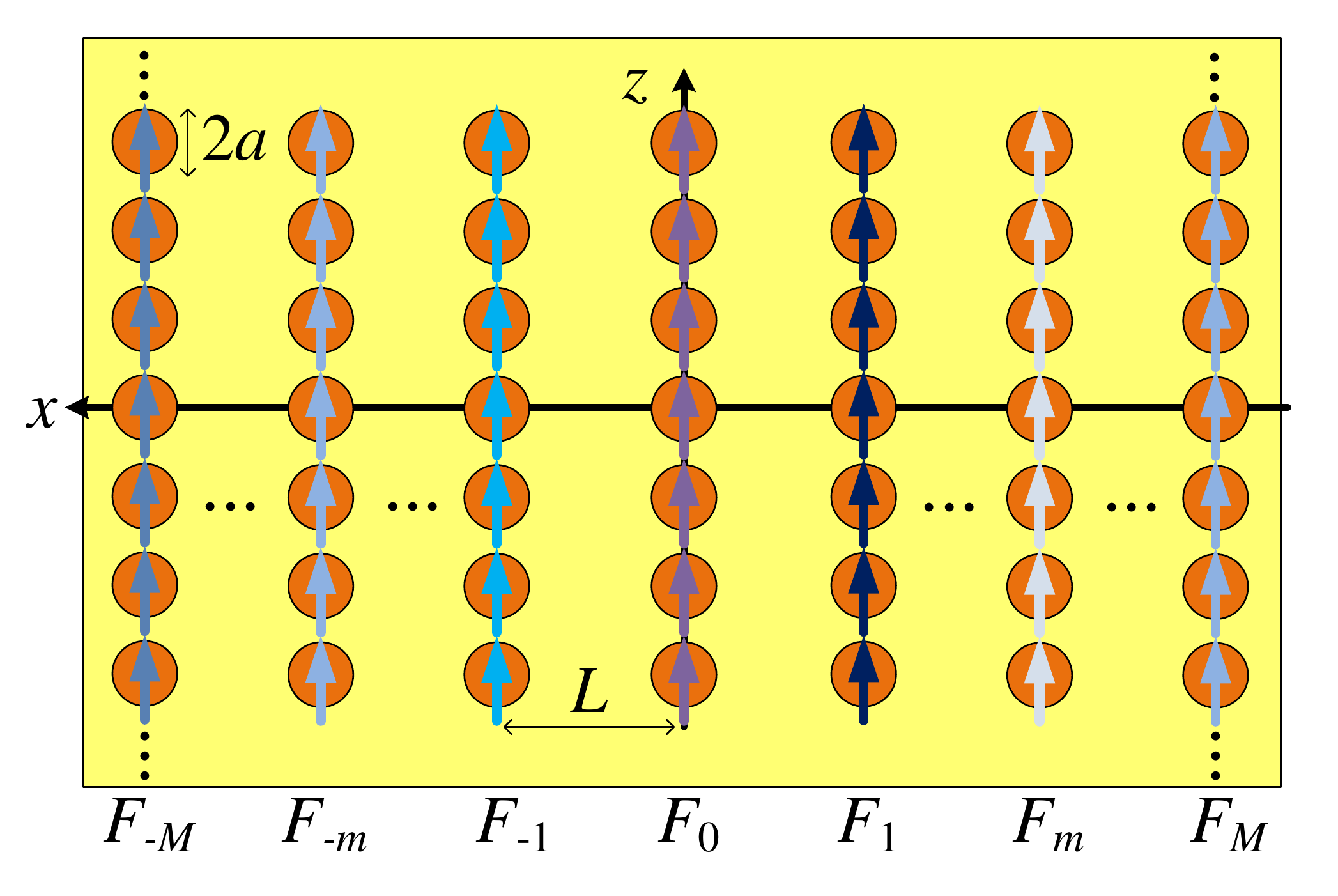}
\caption{Our assumption for invariant fields along $z$ axis can be approximated by considering infinite waveguides terminated at $x=mL$ (and various $z$) which produce exactly the same $z$-polarized electric field $F_m$ for $m=-M,\cdots,M$. Alternatively, one can consider only $(2M+1)$ sources with outputs along the line $(x,y)=(0,0)$ and work at the maximal radiation plane $z=0$.}
\label{fig:Fig2}
\end{figure}

To simplify our analysis, we assume that the phasor of the electric field across the entire zone $\{|x-mL|<a, y=0\}$ equals to $F_m$; such a reduction, renders our problem 2D (field quantities independent from $z$). A possible way to achieve this $z$-independence is proposed in Fig. \ref{fig:Fig2}: along directions parallel to $z$ axis, we spatially repeat infinite times the same waveguides with identical excitation and therefore identical output $z$-polarized electric fields $F_m$. Therefore, our approach works better for laser arrays located across a strip with length $D=2ML$ and arbitrary width $W$; in this way, we create an illusion of $z$ independence which gets more successful for increased $W$. We can assume that the distance between two neighboring couples in the $z$ direction is infinitesimal and thus a single electric component $\hat{\textbf{z}}F_m$ is developed along the axis $(x,y)=(mL, 0)$. In other words, the outputs of the lasers along $x$ axis are uncorrelated each other, while along $y$ direction are the same regardless of the observation point on the imperfect mirror facet. Equivalently, we can say that we are working with a realistic three-dimensional (3D) configuration of $(2M+1)$ lasers whose output ports are located along the line $(z,y)=(0,0)$ and we confine our research to the maximal radiation plane \cite{MaximalRadiationPlane} $z=0$.

The purpose of this inverse analysis is to determine the local field sources $F_m$ with $m=-M,\cdots, M$, which, by radiation into vacuum, produce aggregate electric field with a desired azimuthal pattern $\tilde{G}(\phi)$ in the far region $k_0\rho\rightarrow \infty$. The variables $(\rho, \phi)$ are the polar coordinates which are used interchangeably with $(x,y)$ as indicated in Figs. \ref{fig:Fig1} and \ref{fig:Fig2}. 

\section{From Local Fields to Far-Region Pattern}
\label{Loc2Far}
To quantify the spatial dependence of the electric field $\textbf{E}(x,y)$ within the vacuum half-space $y>0$, we consider the local outputs of the $(2M+1)$ emitters as displacement current sources or, alternatively, as radiating apertures. Furthermore, we can ignore the structure of the cavity lasers array at $y<0$ and replace them by free space; their job was just to provide us with the local fields $F_m$, which have been already taken into account. In this sense, we can work as if $(2M+1)$ sources radiate simultaneously into free-space (for both $y>0$ and $y<0$). We expect that the actual field for $y>0$ will be well-approximated by the aggregate field of the aforementioned sources especially in the far region.

It is straightforward \cite{Balanis, BoundaryValueProblem} to show that the $z$ electric component developed by the $m$-th source alone into free-space is given by:
\begin{eqnarray}
E_m(x,y)=\frac{F_m}{H_0^{(2)}(k_0a)}H_0^{(2)}\left(k_0\sqrt{(x-mL)^2+y^2}\right)
\label{eq:OneSourceField},
\end{eqnarray}
where $k_0$ is the wavenumber of free space and $H_0^{(2)}$ is the Hankel function of zeroth order and second type (solution to the scalar wave equation in cylindrical or polar coordinates \cite{HankelHelmoltz, Balanis}). If one uses the polar coordinate system $(\rho,\phi)$, one finds that the total $z$-polarized field $E=\sum_{m=-M}^ME_m$ of all the considered sources into free space, takes the form:
\begin{eqnarray}
E(\rho,\phi)=\frac{1}{H_0^{(2)}(k_0a)}\sum_{m=-M}^MF_mH_0^{(2)}\left(k_0r A_m(\rho)\right)
\label{eq:AllSourcesField},
\end{eqnarray}
where $A_m(r)=\sqrt{1-2\frac{mL}{\rho}\cos\phi+\left(\frac{mL}{\rho}\right)^2}$. In the far region $(k_0\rho\rightarrow \infty)$, we can find an approximate equivalent if we utilize the asymptotic expansion of Hankel function for large arguments \cite{Stegun}:
\begin{eqnarray}
H_0^{(2)}(k_0R)\sim \sqrt{\frac{2j}{\pi k_0R}}\exp\left(-jk_0R\right),~k_0R\rightarrow\infty
\label{eq:HankelAsymptoticExpansion}.
\end{eqnarray}
In other words, Hankel function far from the source behaves as a cylindrical wave with amplitude vanishing with the square root of the optical distance $k_0R$. If one additionally makes the approximation $A_m(r)\cong 1-\frac{mL}{\rho}\cos\phi$ for the phase in the far region $r\gg ML$, the electric far field produced by the array of lasers at $y>0$ is written as:
\begin{widetext}
\begin{eqnarray}
E(\rho,\phi)\cong\frac{1}{H_0^{(2)}(k_0a)}\exp(-jk_0\rho)\sqrt{\frac{2j}{\pi k_0\rho}}\sum_{m=-M}^MF_m\exp\left(jk_0Lm\cos\phi\right)\equiv
\sqrt{\frac{2j}{\pi k_0\rho}}\exp(-jk_0\rho)G(\phi)
\label{eq:AllSourcesFarField}.
\end{eqnarray}
\end{widetext}
For the amplitude, we perform the less sharp \cite{Balanis} approximation: $A_m(r)\cong 1$ when $k_0\rho\rightarrow \infty$. Also the cylindrical (dimensionless) propagation factor $\sqrt{\frac{2j}{\pi k_0\rho}}e^{-jk_0\rho}$, which is common for the far field of every finite-size source \cite{UnlockingGround}, is dropped for brevity.

Therefore, we have an explicit expression for the azimuthal profile of the far-field $G(\phi)=\sum_{m=-M}^M\frac{F_m}{H_0^{(2)}(k_0a)}e^{jk_0Lm\cos\phi}$ (measured in Volt/meter), produced by $(2M+1)$ infinite series of coupled cavity lasers with local electric fields $F_m$ (measured in Volt/meter) for $m=-M,\cdots, M$.

The aforementioned analysis would be also valid for the 3D problem of a single set of $(2M+1)$ laser emitters located along $x$ axis provided that we do care only about the field distribution on the maximal radiation plane ($z=0$). Each of the $(2M+1)$ waveguides is characterized by the $z$-polarized electric field $F_m$, which is supposed homogeneous into the volume of a small sphere of radius $a$ centralized at $(x,y)=(0,mL)$ and constitutes the end of the corresponding ($m$-th) waveguide ($m=-M,\cdots, M$). For observation points positioned on $xy$ plane, the electric field developed due to the $m$-th laser has a sole $z$ component with complex phasor $E_m(\rho,\phi)=F_m\frac{k_0a}{e^{-jk_0a}}\frac{e^{-jk_0\rho A_m(\rho)}}{k_0\rho A_m(\rho)}$. Note that, since we are referring to the slice $z=0$ of our 3D space, the distance of an observation point from $z$ axis (cylindrical radial coordinate $\rho$) is the same with its distance from the origin (spherical radial coordinate).

In this way, one can obtain an expression for the aggregate far field almost the same to (\ref{eq:AllSourcesFarField}): $E(\rho,\phi)\cong \frac{k_0a}{e^{-jk_0a}}\frac{e^{-jk_0\rho}}{k_0r}\sum_{m=-M}^MF_m e^{jk_0Lm\cos\phi}$. In fact, we realize that the only difference from (\ref{eq:AllSourcesFarField}) is the $r$-dependence, describing the field of a spherical source, which expresses the inevitable attenuation because of distribution of power around larger and larger semicircles of radius $r$. However, this common factor $\frac{e^{-jk_0\rho}}{k_0r}$ does not participate in our inverse problem; we do count only the $\phi$-dependent pattern $G(\phi)=\frac{k_0a}{e^{-jk_0a}}\sum_{m=-M}^MF_m e^{jk_0Lm\cos\phi}$. Therefore, the approach of making the variational sum $G(\phi)$ as similar as possible to the target pattern $\tilde{G}(\phi)$, which is the objective of the next Section \ref{DesFar2Opt}, remains unaltered either we consider the 2D problem or that special case of 3D one.

\section{From Desired Far-Region Pattern to Optimal Local Fields}
\label{DesFar2Opt}
Ideally, we should aim at achieving the exact equality between the obtained field pattern $G(\phi)$ created by the lasers and the target (desired) pattern $\tilde{G}(\phi)$ for all the angles $0<\phi<180^{\circ}$ of the upper half space $y>0$. However, this is not always possible and in order to develop a general technique which can be applied to any variation of $\tilde{G}(\phi)$, we should search for the optimal set of output fields $\left\{F_m,~m=-M,\cdots, M\right\}$ mimicking best the desired pattern.

We notice that $G(\phi)$ is expressed as a finite sum of the basis functions set: $\left\{p_m(\phi)=e^{jk_0Lm\cos\phi},~m=-M,\cdots, M\right\}$. These functions do not constitute a complete set due to the finite size $(2M+1)$ of their population. Furthermore, the variations of $p_m(\phi)$ are dependent on the parameter $k_0L$, which expresses how sparse is our emitter lattice. Choosing a small $k_0L$, for fixed $M$, will make us \textit{lose the ability to imitate rapid $\phi$-variations}; on the other hand, choosing a large $k_0L$ will \textit{decrease our ``resolution'' of mimicking patterns} since only specific waveforms, very different each other, would be possible to get produced. Therefore, a careful study of the degrees of azimuthal variational freedom and numerical robustness of our method with respect to $k_0L$ is required.

As far as the orthogonality of the basis functions set $\left\{p_m(\phi),~m=-M,\cdots, M\right\}$ is concerned, it does not exist either (same to the completeness property). In particular, their cross product over the azimuthal range $0<\phi<180^{\circ}$ is nonzero; however, it can be analytically \cite{Stegun} evaluated:
\begin{widetext}
\begin{eqnarray}
\frac{1}{\pi}\int_0^\pi p_m(\phi)p_n^*(\phi)d\phi=
\frac{1}{\pi}\int_0^\pi\exp\left(jk_0L(m-n)\cos\phi\right)d\phi=
J_0(k_0L(m-n))\equiv S_{nm}
\label{eq:CrossProduct},
\end{eqnarray}
\end{widetext}
where $m,n=-M,\cdots, M$ and $J_0$ is the Bessel function of zeroth order. If we try to project the ideal equality $G(\phi)=\tilde{G}(\phi)$ on the conjugate $(*)$ set of basis functions, namely adopt the Galerkin \cite{Galerkin} approach (where the testing functions are complex conjugate of the basis functions), we obtain the $(2M+1)\times(2M+1)$ linear system $\textbf{S}\cdot \textbf{f}=\textbf{v}$ with unknown the optimal vector of fields $\textbf{f}=\left[F_{-M} \cdots F_{M}\right]^T$. The matrix of the system $\textbf{S}=[S_{nm}]$ solely contains values of function $J_0$ as indicated by (\ref{eq:CrossProduct}), which appears in a number of intriguing physical situations \cite{EliotMontrol, Kovanis1996_PRA}. Note that $\textbf{S}$ is real, symmetric and its diagonal elements are all equal to one: $J_0(0)=1$. When it comes to the constant vector $\textbf{v}=\left[V_{-M} \cdots V_{M}\right]^T$, it is defined by the following general term:
\begin{eqnarray}
V_n=\frac{H_0^{(2)}(k_0a)}{\pi}\int_0^{\pi}\tilde{G}(\phi)\exp\left(-jk_0Ln\cos\phi\right)d\phi
\label{eq:ConstantVector},
\end{eqnarray}
which are the components of the ideal pattern $\tilde{G}(\phi)$ in the base of testing functions set: $\left\{p_n^*(\phi)=e^{-jk_0Ln\cos\phi},~n=-M,\cdots, M\right\}$. 

\begin{figure}[htbp]
\centering
\subfigure[]{\includegraphics[width=6.8cm]{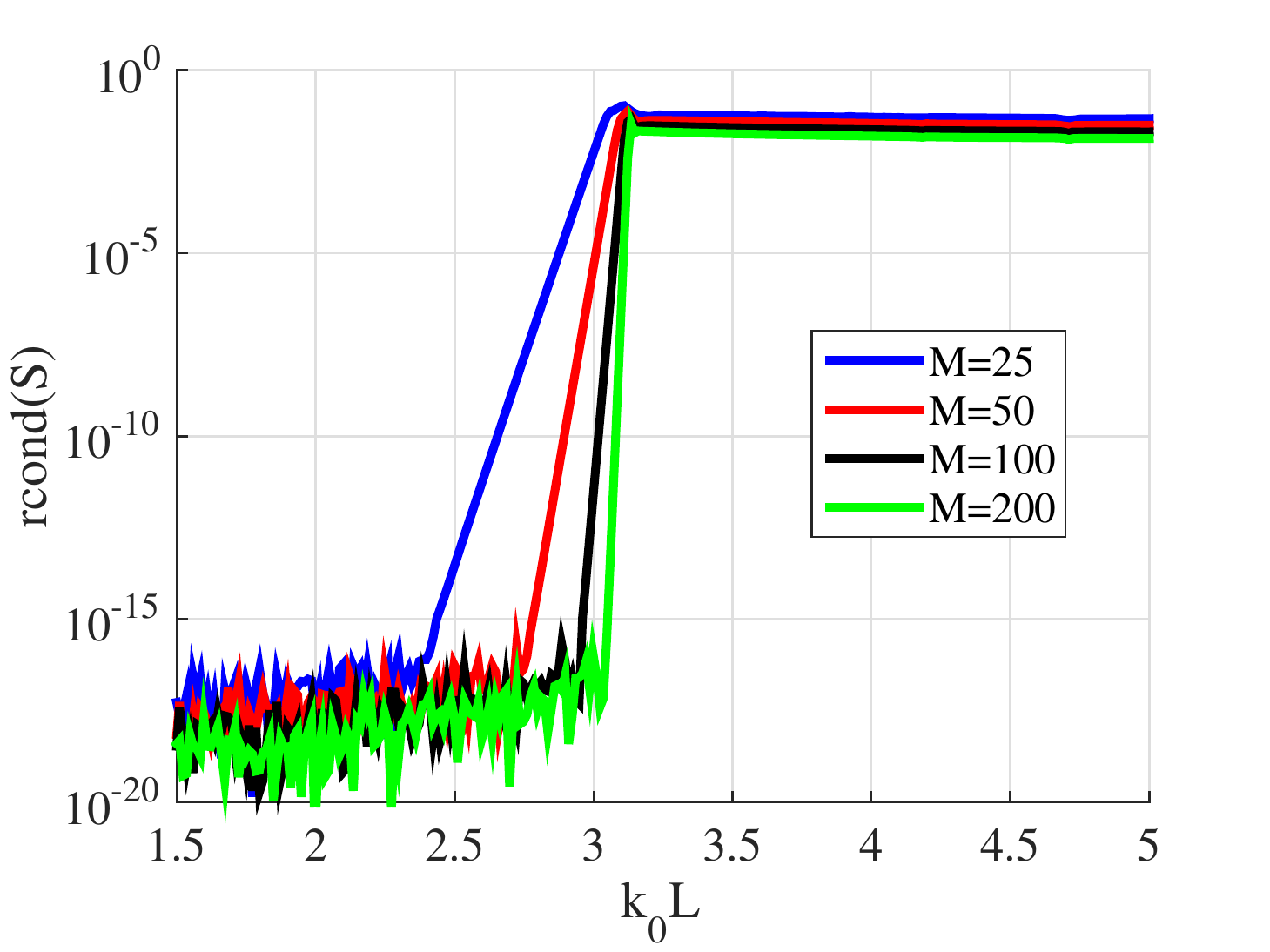}
   \label{fig:Fig4a}}
\subfigure[]{\includegraphics[width=6.8cm]{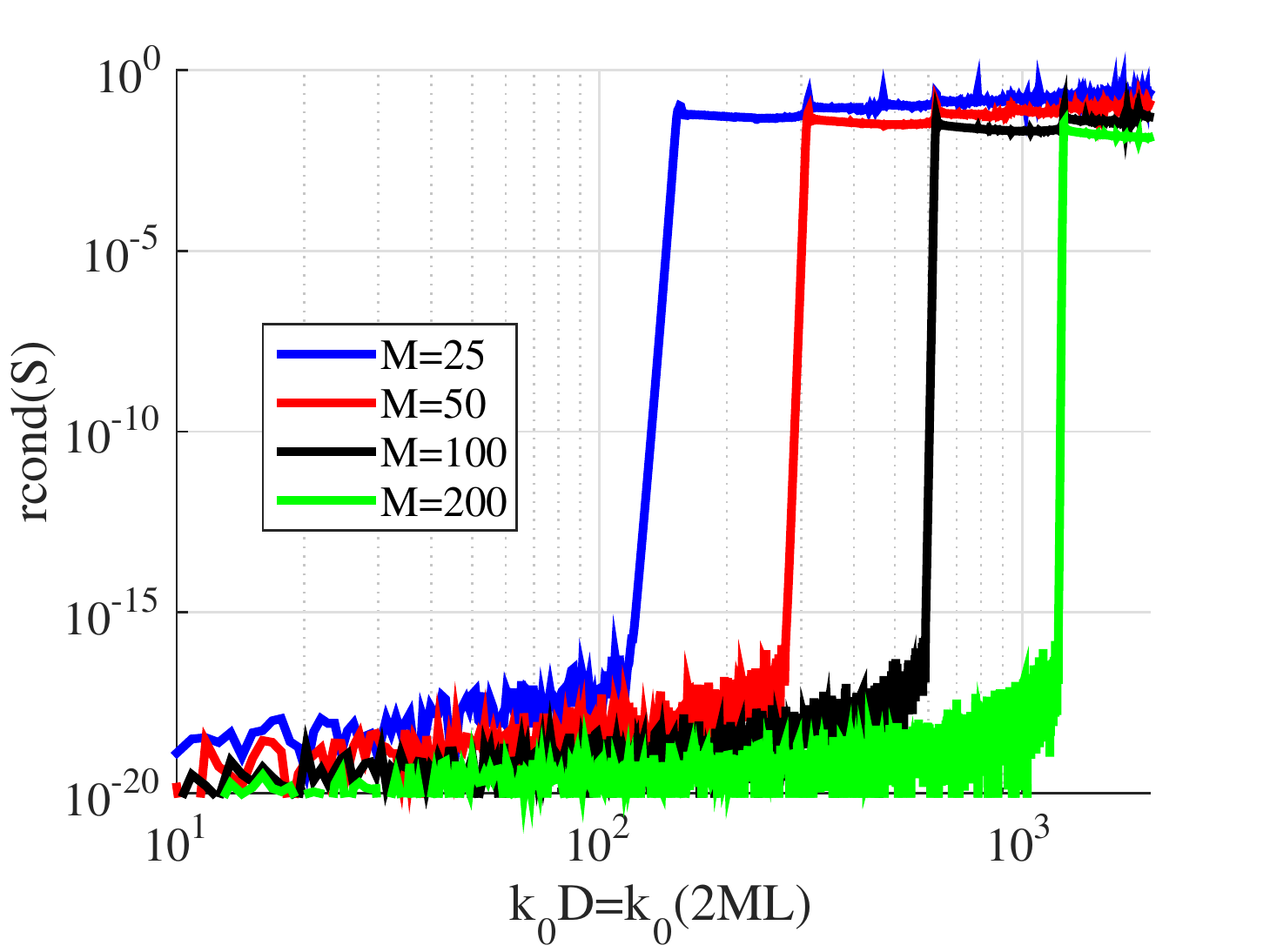}
   \label{fig:Fig4b}}
\caption{The reciprocal condition number of our linear system matrix $\textbf{S}$ as function of : (a) the optical period of our laser lattice $k_0L$ and (b) the optical size of our entire cluster of lasers $k_0D=k_0(2ML)$, for several numbers of lasers $M$.}
\label{fig:Figs4}
\end{figure}    

Since the numerical behavior of the matrix $\textbf{S}$ determines the accuracy and robustness of the linear system with respect to $\textbf{f}$, let us examine the reciprocal condition number ${\rm rcond}$ of $\textbf{S}$. Such a number \cite{ReciprocalCondition} returns an estimate for the reciprocal condition of  $\textbf{S}$ in 1-norm ($\textbf{S}$ is well-conditioned for ${\rm rcond}$$(\textbf{S})$$\rightarrow 1^-$ and $\textbf{S}$ is badly-scaled for ${\rm rcond}$$(\textbf{S})$$\rightarrow 0^+$). In Fig. \ref{fig:Fig4a}, we show $0<{\rm rcond}(\textbf{S})<1$ as function of the optical period of the laser lattice $k_0L$ for various populations $M$ of lasers. The curves corresponding to $M=25$ and $M=50$ are shown with blue (dark gray) and red (lighter gray) color respectively, while the data for $M=100$ and $M=200$ are depicted in black and green (light gray) color respectively. One clearly notices that our numerical inversion is more accurate for larger $k_0L$. This conclusion may be anticipated since the variation of $J_0(k_0R)$ resembles a cosine with decaying amplitude (as $1/\sqrt{|k_0R|}$ \cite{Stegun} from (\ref{eq:HankelAsymptoticExpansion})); therefore, for a high $k_0L$ all the off-diagonal elements of the matrix $\textbf{S}$ would be small compared to its unitary diagonal ones and $\textbf{S}$ would behave robustly as a close-to-diagonal matrix. It should be stressed that the number of lasers $M$ does not play a crucial role in the numerical behavior of $\textbf{S}$; in all cases, regardless of $M$, matrix $\textbf{S}$ acquires a good condition approximately for $k_0L>3$. By increasing the number of lasers, one just makes the overall device longer. However, in Fig. \ref{fig:Fig4b}, where ${\rm rcond}$$(\textbf{S})$ is represented as function of the total size of our laser cluster $k_0D=k_0(2ML)$, we can apparently observe that sparser structures (smaller $M/k_0D$) exhibit better numerical behavior.

But just utilizing a long array with large optical distances $k_0L$ between two neighboring lasers would be the correct choice? For sure, based on Figs \ref{fig:Figs4}, will give a very low error of the linear system $||\textbf{S}\cdot\textbf{f}-\textbf{v}||$, namely will return fields $\textbf{f}$ satisfying exactly the projection equations. But what is happening with the actual error $\int_0^{\pi}|\tilde{G}(\phi)-G(\phi)|d\phi$ between the target pattern $\tilde{G}(\phi)$ and the produced far-field $G(\phi)$ which is the crucial quantity determining the degree of success of our approach? As indicated above, a large $k_0L$ will substantially demote the ability of the variational finite sum $G(\phi)$ to mimic the desired response $\tilde{G}(\phi)$. For this reason, we should balance between the demand for high $k_0L$ ensuring stable numerical behavior (small linear system error) and the requirement of low $k_0L$ (accompanied with suitably large $M$) leading to significant capacity for $\phi$-variation adjustment (small deviation between the actual $G$ and ideal target $\tilde{G}$ pattern).

To test the efficiency of our process, let us consider the following family of desired patterns $\tilde{G}(\phi)$:
\begin{eqnarray}
\tilde{G}(\phi)=\exp(-\beta\phi)\left[1+A\cos(\alpha\phi)\right],
\label{eq:DesiredPatternFamily}
\end{eqnarray}
where the magnitude $A$ expresses the difference of $\tilde{G}(\phi)$ from an omni-directional pattern, the number $\alpha$ determines the rapidness of oscillations with respect to $\phi$ and the quantity $\beta$ specifies the envelope trend. It is a demanding and ``unnatural'' formula which contains several types of azimuthal variations; if our method performs well with such a not easily satisfied target, it will have good performance with more natural beam shapes.

\begin{figure}[htbp]
\centering
\subfigure[]{\includegraphics[width=4.2cm]{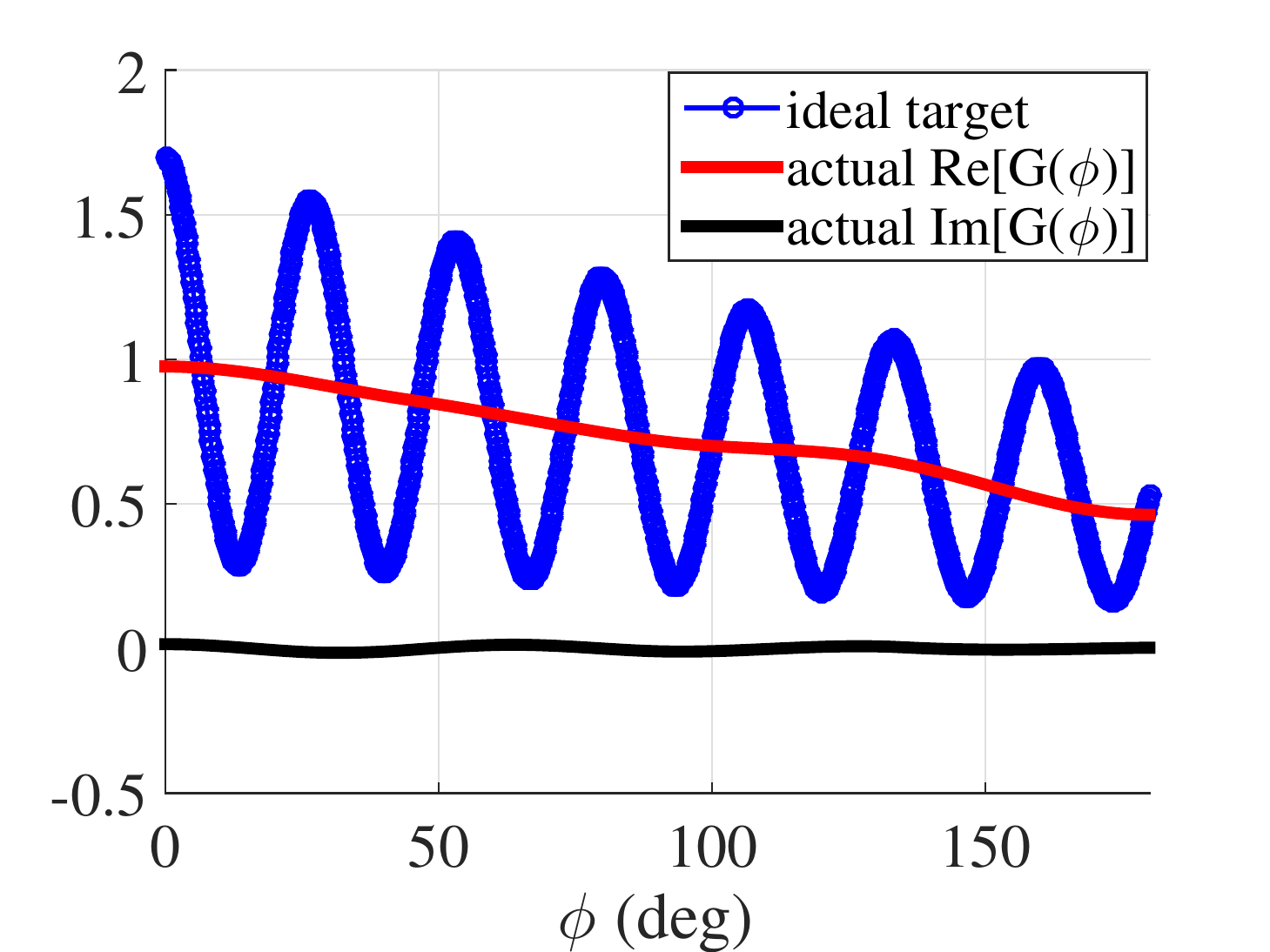}
   \label{fig:Fig6a}}
\subfigure[]{\includegraphics[width=4.2cm]{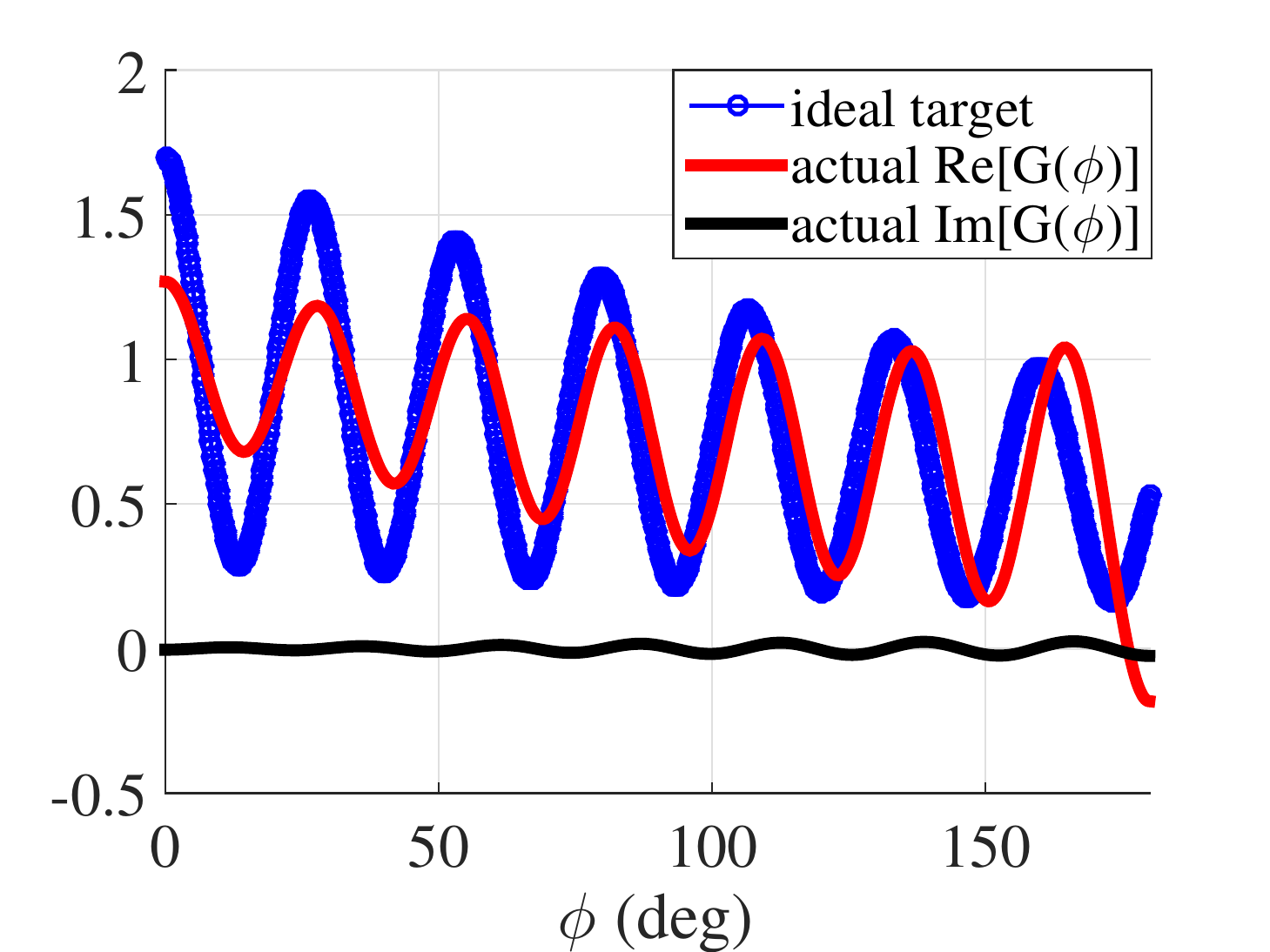}
   \label{fig:Fig6b}}
\subfigure[]{\includegraphics[width=4.2cm]{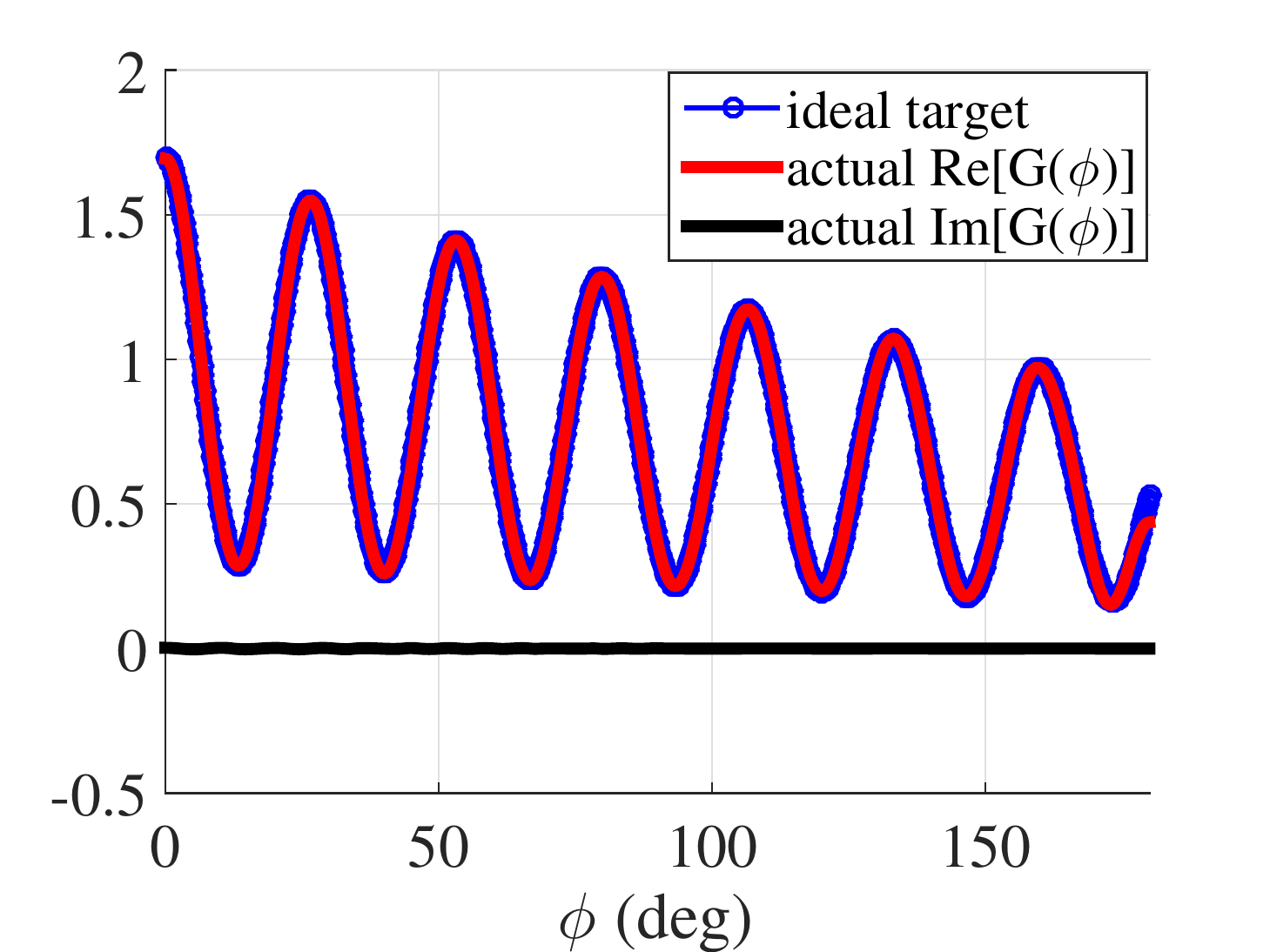}
   \label{fig:Fig6c}}
\subfigure[]{\includegraphics[width=4.2cm]{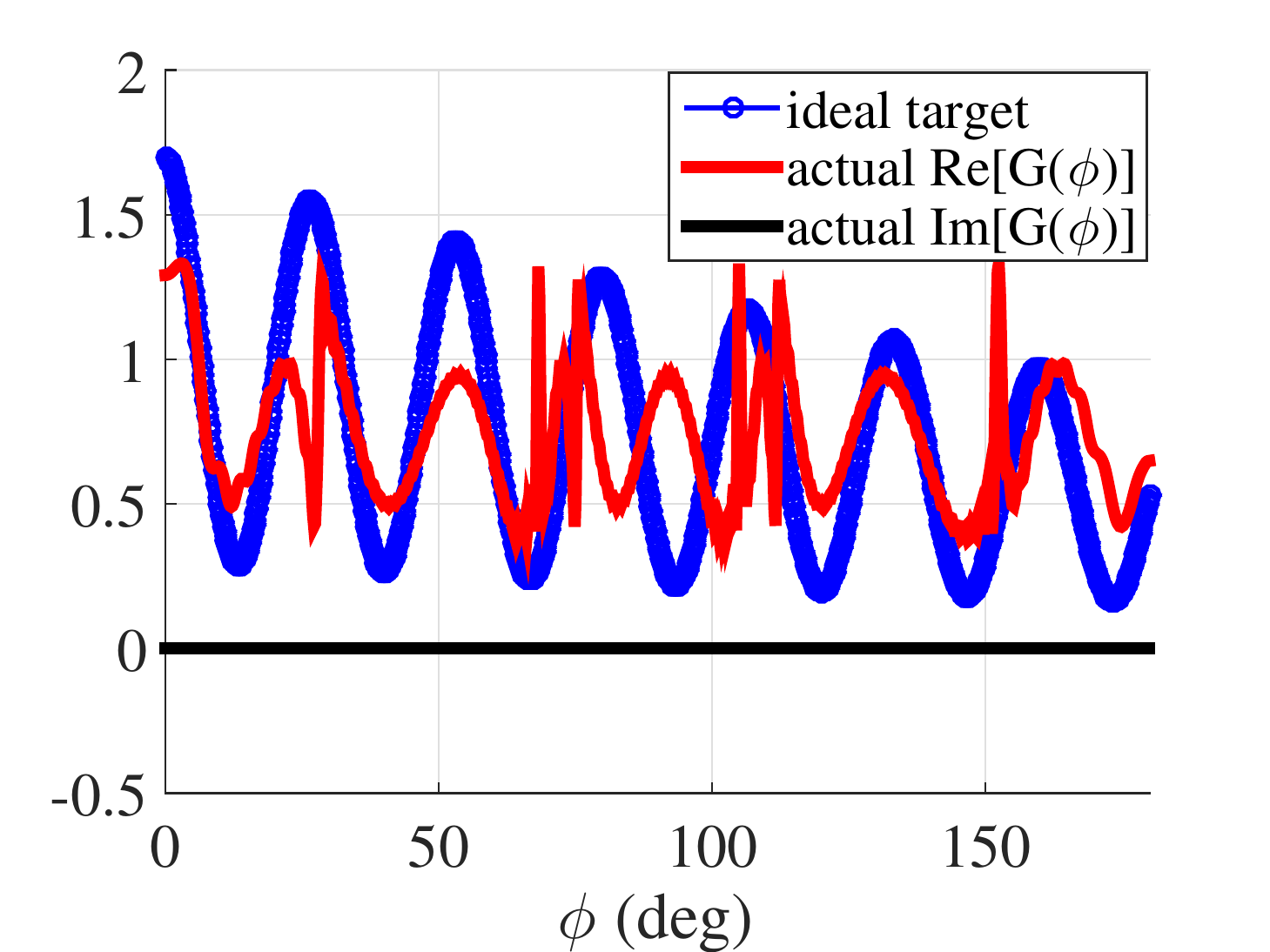}
   \label{fig:Fig6d}}
\caption{The ideal target $\tilde{G}(\phi)$ and the optimal actual pattern $G(\phi)$ (both real and imaginary parts) as functions of azimuthal angle $\phi$ for: (a) $k_0L=0.005$, (b) $k_0L=0.1$, (c) $k_0L=1$, (d) $k_0L=10$. Plot parameters: $A=0.7$, $\alpha=13.5$, $\beta=0.2$, referring to (\ref{eq:DesiredPatternFamily}) and $M=50$.}
\label{fig:Figs6}
\end{figure}

In Figs \ref{fig:Figs6}, we represent with solid red (light gray) line an ideal target $\tilde{G}(\phi)$ from (\ref{eq:DesiredPatternFamily}) with $A=0.7$, $\alpha=13.5$, $\beta=0.2$ for $M=50$ and we apply the technique described above to obtain actual patterns $G(\phi)$ for various optical distances between the emitters $k_0L$. The real part of $G(\phi)$ is shown as thick curves made of consecutive circles, while the imaginary part has usually small magnitude and is sketched with thin black curves close to horizontal $\phi$ axis. In Fig. \ref{fig:Fig6a}, where a very tiny $k_0L$ is selected, our method totally fails: the real parts differ substantially. Such a poor result is not due to the fact that the matrix $\textbf{S}$ of the system is badly-conditioned but mainly because the describing set of functions $\left\{p_m(\phi)=e^{jk_0Lm\cos\phi},~m=-M,\cdots, M\right\}$ does not have components with $\phi$-variation rapid enough to mimic the waveform $\tilde{G}(\phi)$. Indeed, the maximally varying (with respect to $\phi$) basis function from the aforementioned set is the one corresponding to $m=M$: $p_M(\phi)=e^{jk_0LM\cos\phi}=e^{jk_0D\cos\phi/2}$. The most rapid azimuthal variation of this function happens around $\phi=0^{\circ}$ and $\phi=180^{\circ}$ since there the argument of the harmonic exponential gets maximized. In the vicinity of such angles, we obtain $|\cos\phi|\cong 1$. Therefore, the degree $u$ of the largest \textit{significant} harmonic $e^{+ju\phi}$ contained in $p_M(\phi)$ (and accordingly into the entire set of basis functions) is approximately the $u=\lfloor k_0D/2 \rfloor=M\lfloor k_0L \rfloor$. In other words, if we have a desired target with rapid azimuthal variation dictated by a maximum significant harmonic $u_{max}>u=M \lfloor k_0L \rfloor$, then, it is not possible for the set of functions $\left\{p_m(\phi)=e^{jk_0Lm\cos\phi},~m=-M,\cdots, M\right\}$ to capture its waveform. This is the case of Fig. \ref{fig:Fig6a}.

If we select a higher $k_0L$ as in Fig. {\ref{fig:Fig6b}}, again the obtained pattern $G$ is substantially different from the desired $\tilde{G}$ and also a significant imaginary part for $G(\phi)$ is obtained (instead of zero). Once more, the difference is not attributed to the numerical inversion as an outcome of an unstable matrix $\textbf{S}$ since the error $||\textbf{S}\cdot \textbf{f}-\textbf{v}||$ is negligible; the problem is again that $k_0L<u_{max}/M$ as in Fig. \ref{fig:Fig6a}. Therefore, we realize that the small ${\rm rcond}$$(\textbf{S})$ indicated in Figs \ref{fig:Figs4}, is not an issue; unless we are talking for an extremely badly conditioned matrix, it can be numerically inverted with negligible numerical error in MATLAB\textsuperscript{\textregistered} environment. Despite the fact that a warning message is appeared at the MATLAB\textsuperscript{\textregistered} command line (automatically activated when ${\rm rcond}$$(\textbf{S})$ falls below a specific threshold), the computational platform has sophisticated toolboxes with adaptive algorithms and transformations which can handle well the cases of non-robust systems $\textbf{S}\cdot \textbf{f}=\textbf{v}$. The symmetric nature of our matrix $\textbf{S}$ is definitely contributing towards this direction \cite{LinearAlgebra, MATLABRef, MATLABRef2}.

Contrary to Figs \ref{fig:Fig6a} and \ref{fig:Fig6b}, in \ref{fig:Fig6c}, the described method works perfectly and the variation of $\tilde{G}(\phi)$ is almost flawlessly captured with negligible error (in both real and imaginary part). However, as it is obvious from Fig. \ref{fig:Fig4a}, the matrix of the system $\textbf{S}$ continues to be close-to-singular; another indication that it does not damage the efficiency of the proposed technique. Of course, such a badly scaled matrix will have an impact on the magnitude of the unknown vector $\textbf{f}$, namely the field amplitudes $\left\{F_m,~m=-M,\cdots, M\right\}$; nevertheless, we do not care much about it as we seek for fields and waveforms normalized by their maximal value (relative results). 

In Fig. \ref{fig:Fig6d}, we increase further the optical period of the array of our lasers and we surprisingly realize that our method fails too. The reason this time is the low resolution potential of the describing set of functions originating from the fact that its members are very different each other which means that they can mimic successfully only few isolated patterns (not a continuous area in pattern space). Empirically, we notice that \textit{such a harming effect happens when the matrix} $\textbf{S}$ \textit{is well-scaled!} To put it alternatively, a linear system very numerically robust (with large ${\rm rcond}$$(\textbf{S})$) is accompanied by a totally inefficient set of basis function (substantial difference between actual pattern $G(\phi)$ and target pattern $\tilde{G}(\phi)$. To avoid such a failure, we should select $k_0L$ somehow less than 3 as indicated by Fig. \ref{fig:Fig4a}. 

If one combines the two aforementioned restrictions for $k_0L$, one can obtain the following rule of thumb:
\begin{eqnarray}
\frac{u_{max}}{M}\lesssim k_0L \lesssim 3
\label{eq:RuleOfThumb},
\end{eqnarray}
for a fixed number of  $(2M+1)$ available lasers and a desired pattern $\tilde{G}(\phi)$ with maximal significant azimuthal component $e^{\pm ju_{max}\phi}$. If $u_{max}/M>3\Rightarrow M<3u_{max}$, then \textit{we do not have enough lasers to mimic such a rapidly oscillating far field}. Inequality (\ref{eq:RuleOfThumb}) is the key mathematical conclusion of this work.

\begin{figure}[htbp]
\centering
\subfigure[]{\includegraphics[width=4.2cm]{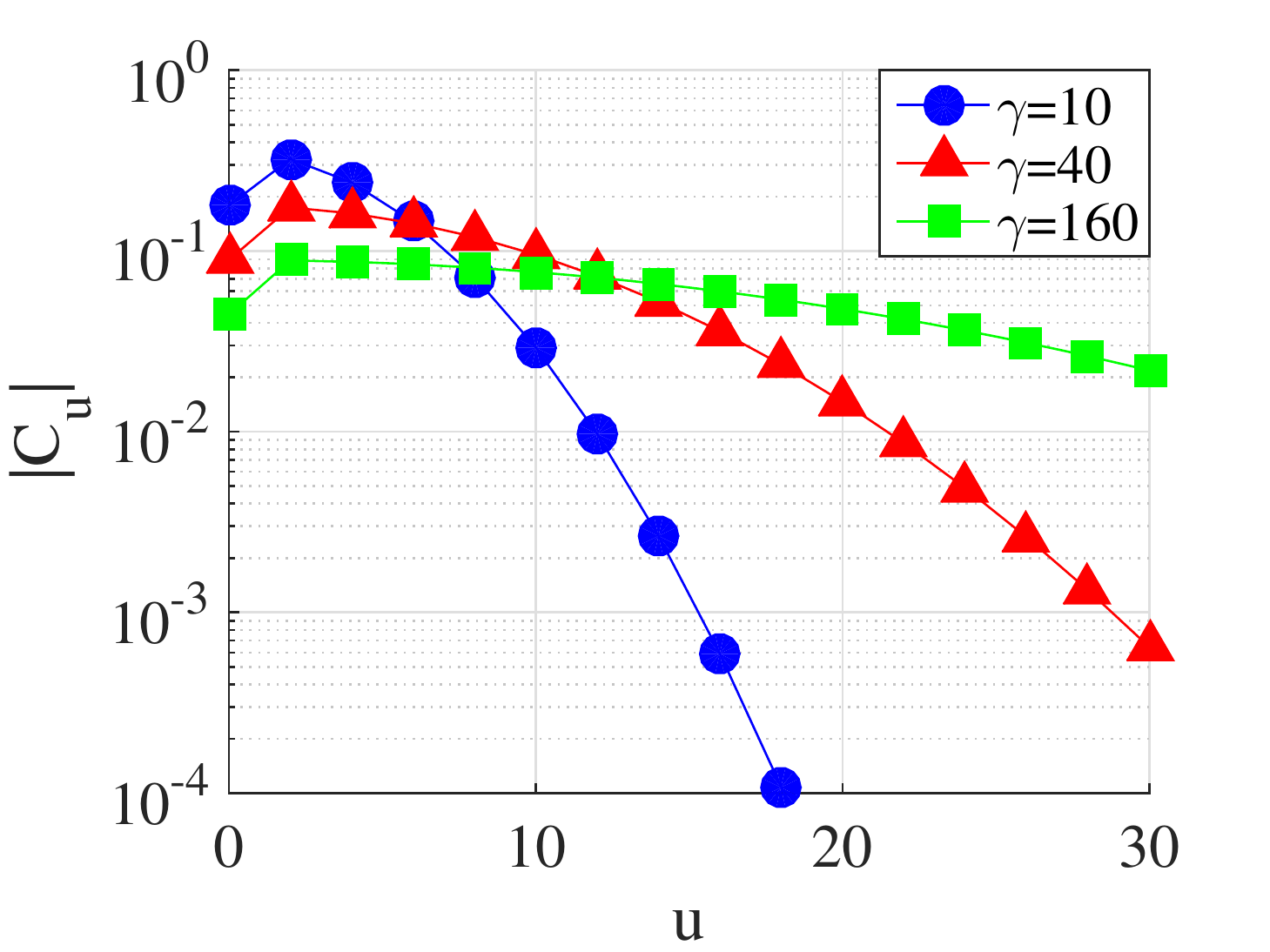}
   \label{fig:Fig7a}}
\subfigure[]{\includegraphics[width=4.2cm]{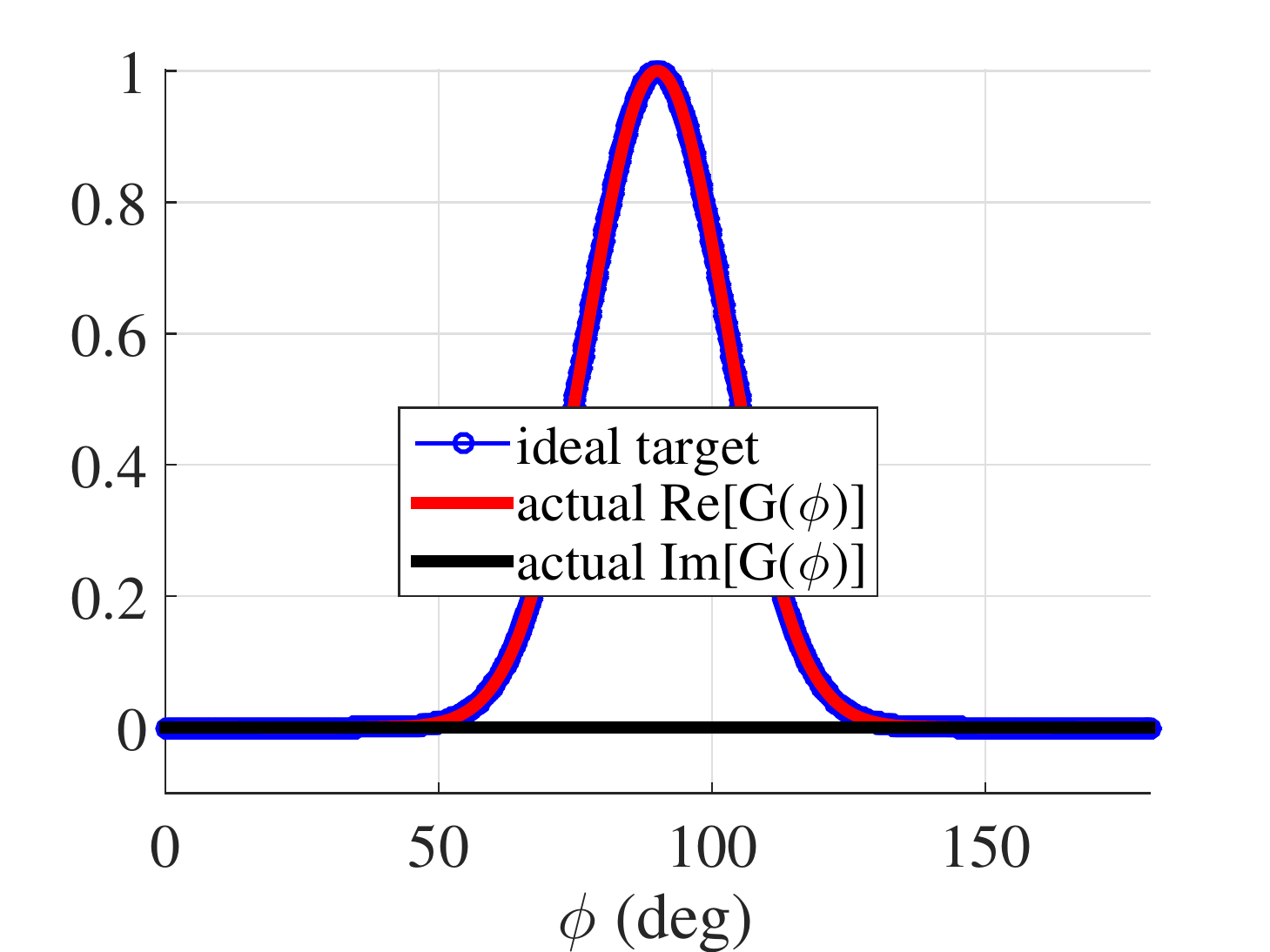}
   \label{fig:Fig7b}}
\subfigure[]{\includegraphics[width=4.2cm]{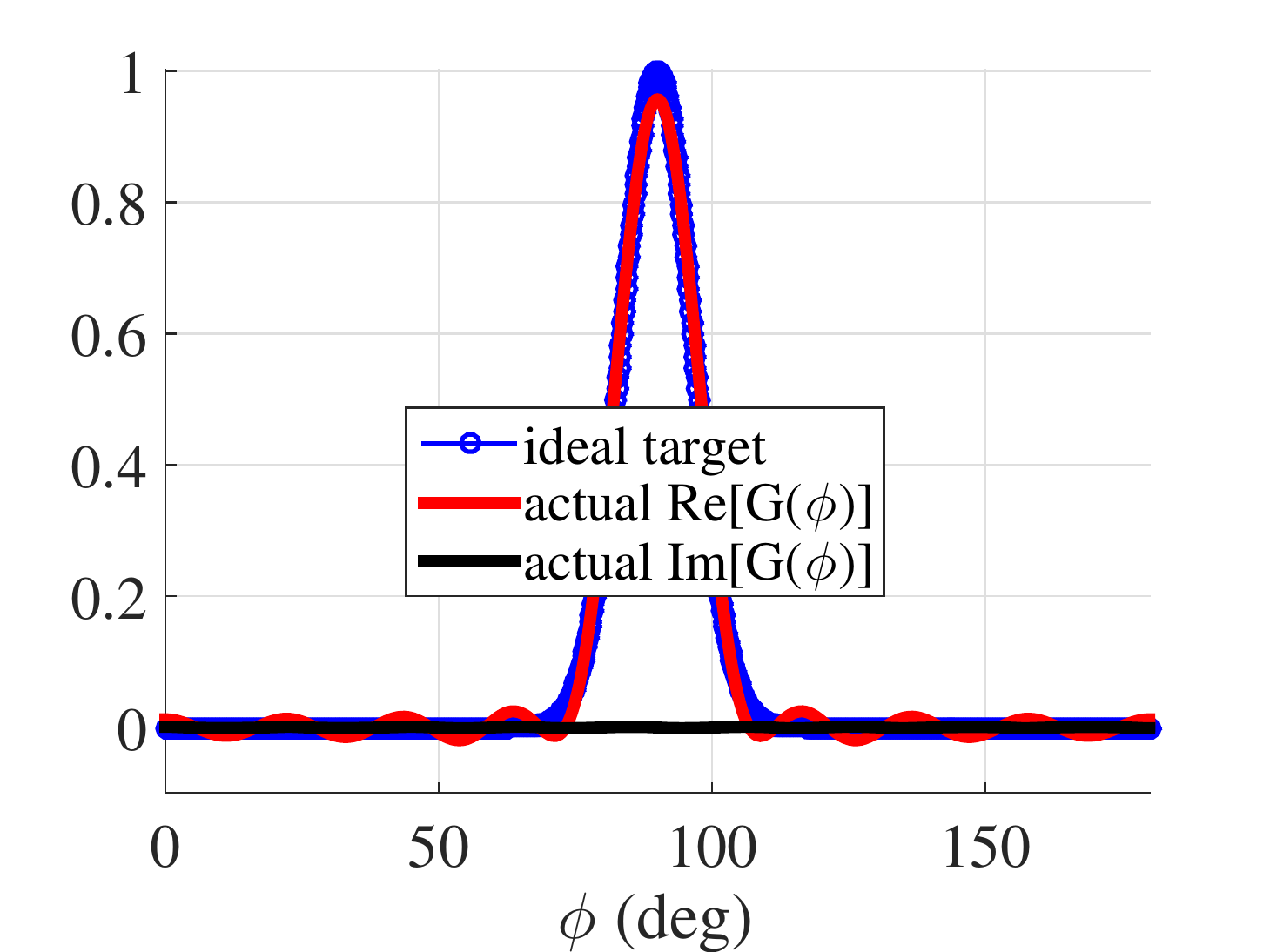}
   \label{fig:Fig7c}}
\subfigure[]{\includegraphics[width=4.2cm]{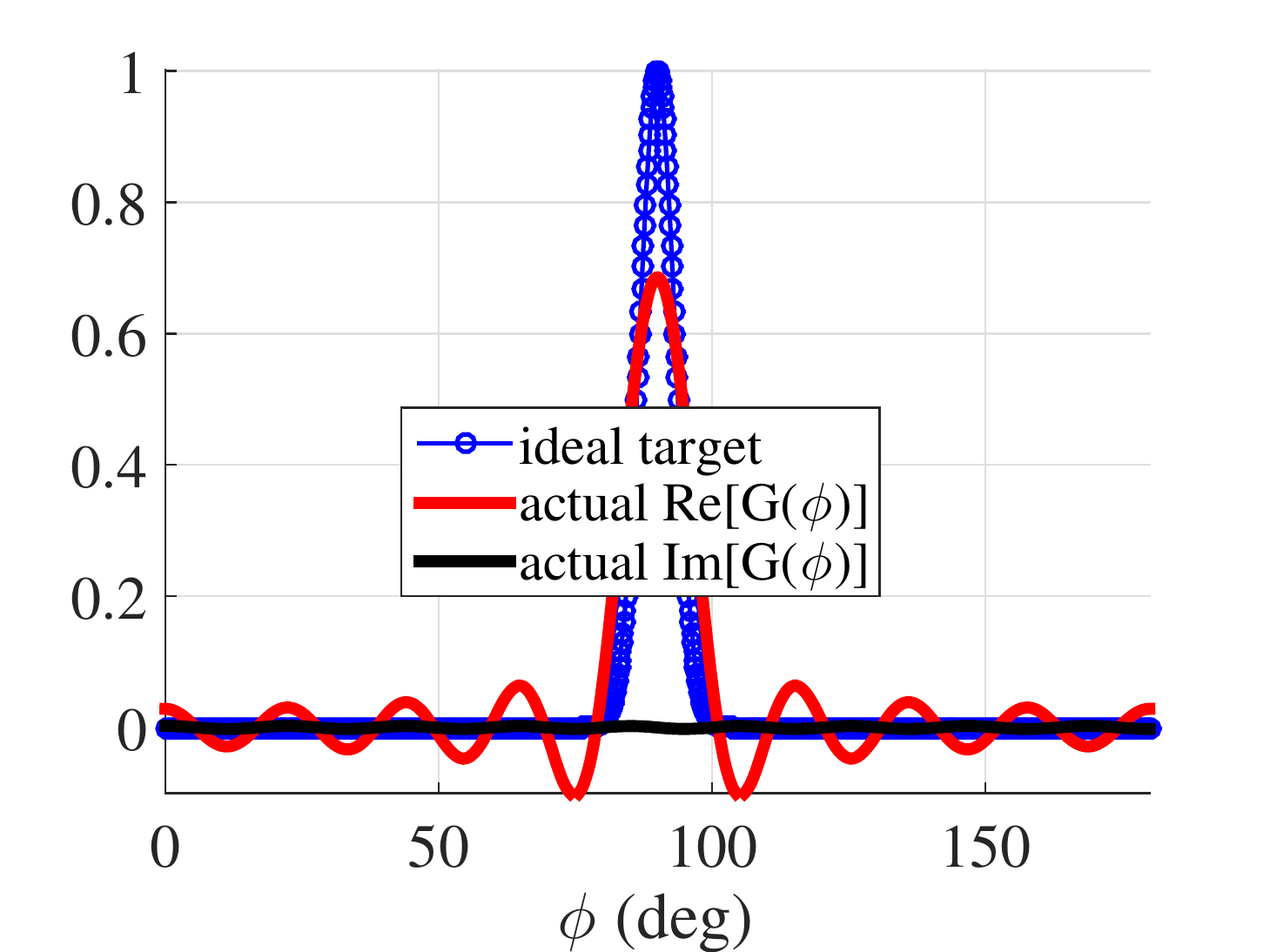}
   \label{fig:Fig7d}}
\caption{(a) The magnitude of the Fourier coefficients $|C_u|$ as function of the azimuthal order $u$ for various target patterns, referring to (\ref{eq:DesiredPatternFamily2}). Continuous curves are guides to the eye only and odd-ordered coefficients are identically zero. The ideal target $\tilde{G}(\phi)$ and the optimal actual pattern $G(\phi)$ (both real and imaginary parts) are shown as functions of azimuthal angle $\phi$ for: (b) $\gamma=10$, (c) $\gamma=40$, (d) $\gamma=160$. Plot parameters: $\vartheta=90^{\circ}$, referring to (\ref{eq:DesiredPatternFamily2}), $k_0L=0.1$ and $M=80$.}
\label{fig:Figs7}
\end{figure}

The elegant, though empirical and approximate, double inequality (\ref{eq:RuleOfThumb}), can be additionally verified if we pose a more realistic target pattern describing tilted directive beams. In particular, let us consider:
\begin{eqnarray}
\tilde{G}(\phi)=\exp\left(-\gamma(\phi-\vartheta)^2\right)
\label{eq:DesiredPatternFamily2},
\end{eqnarray}
where $\gamma>0$ determines how directive would be the main lobe of the radiation pattern around angle $\phi=\vartheta$. As in the target pattern of (\ref{eq:DesiredPatternFamily}), 2$\pi$-periodicity is not required since the regarded area $(0<\phi<180^{\circ})$ is not a complete circle. We can directly express the (even expansion of ) $\tilde{G}$ as Fourier series; namely: $\tilde{G}(\phi)=\sum_{u=0}^{+\infty}C_u\cos(u\phi)$. In order to estimate the maximal significant azimuthal order $u_{max}$, we represent (Fig. \ref{fig:Fig7a}) the magnitudes of the Fourier coefficients $|C_u|$ as function of $u$ for various parameters $\gamma$ of (\ref{eq:DesiredPatternFamily2}), while we keep constant the maximal radiation at $\theta=90^{\circ}$. It is clear that $C_u=0$ for odd $u$ and the continuous curves are guides to the eye only. Given the fact that a substantial $\gamma$ corresponds to narrow beams, it is natural that more Fourier coefficients are required in order to approximate the waveforms at a satisfying degree; therefore, $u_{max}$ is increasing with $\gamma$. In Figs \ref{fig:Fig7b}-\ref{fig:Fig7d}, we try to mimic the far-field response for each of the aforementioned three cases with a specific set of lasers $(M=80)$ and fixed placement $(k_0L=0.1)$. The same coloring rules as in Figs \ref{fig:Figs6} apply. It is obvious that the optimal output fields as calculated by the proposed method make an excellent reproduction of $\tilde{G}(\phi)$ for $\gamma=10$ (Fig. \ref{fig:Fig7b}). However, a small error between $\tilde{G}(\phi)$ and $G(\phi)$ is recorded for the more directive beam of $\gamma=40$ (Fig. \ref{fig:Fig7c}) whose azimuthal variation is more abrupt. Finally, our approach does not perform well for the steepest beam case (Fig. \ref{fig:Fig7d}), where spurious sidelobes are appeared and the main maximum is substantially weakened. We can draw the conclusion that since $u_{max}$ is the only one that changes in Figs \ref{fig:Fig7b}-\ref{fig:Fig7d}, we have a $u_{max}<Mk_0L$ in the first two cases ($\gamma=10, 40$ in Figs \ref{fig:Fig7b}, \ref{fig:Fig7c}) and $u_{max}>Mk_0L$ in the last one ($\gamma=160$ in Fig. \ref{fig:Fig7d}). It is important to stress that these conditions are approximate in the same degree that the maximum significant azimuthal order $u_{max}$ and both inequalities of rule of thumb (\ref{eq:RuleOfThumb}) are also roughly evaluated. That is why we expressed the double inequality (\ref{eq:RuleOfThumb}) in terms of \textit{greater/less than about} symbols.
 
\begin{figure}[htbp]
\centering
\subfigure[]{\includegraphics[width=4.2cm]{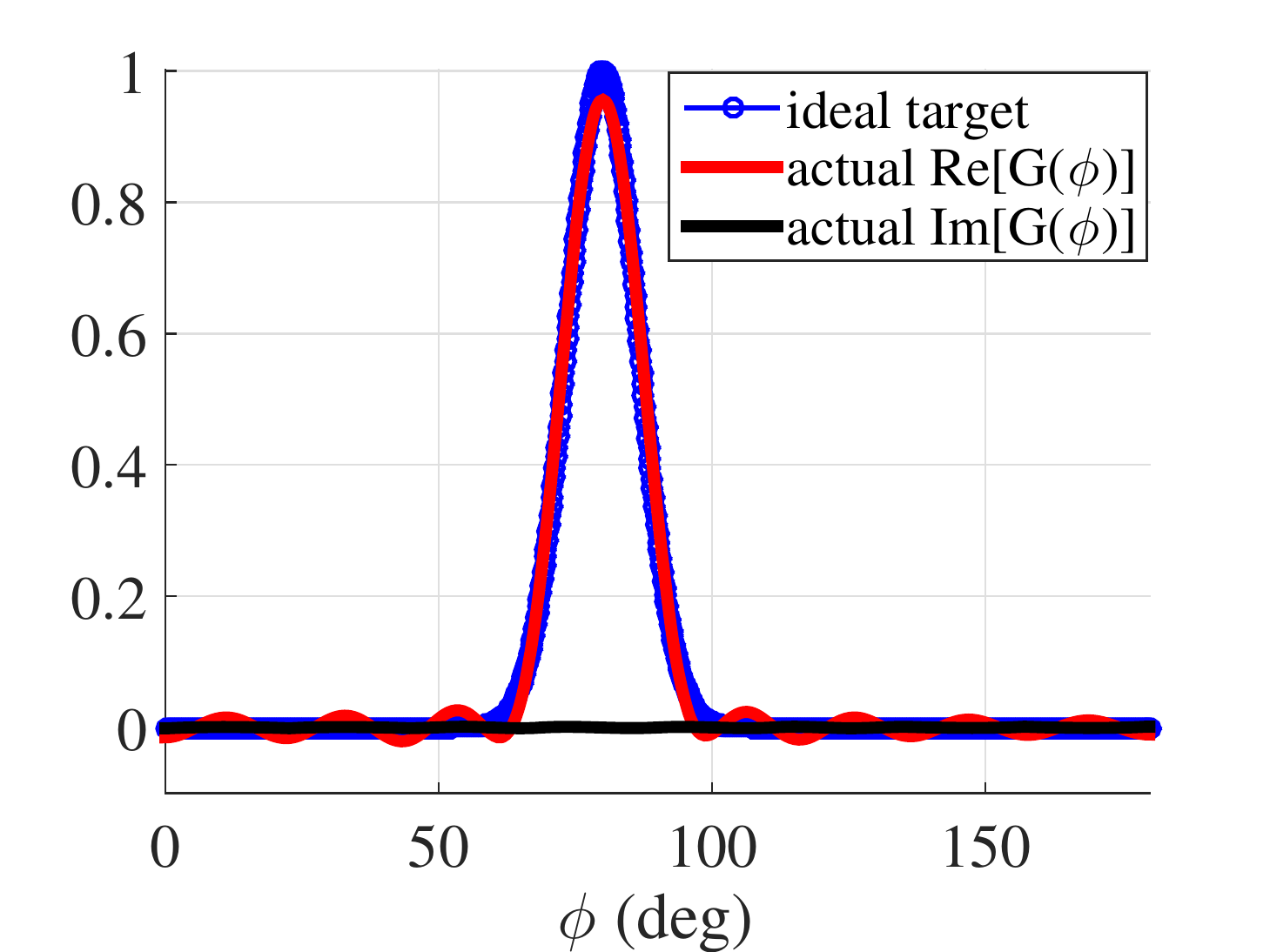}
   \label{fig:Fig8a}}
\subfigure[]{\includegraphics[width=4.2cm]{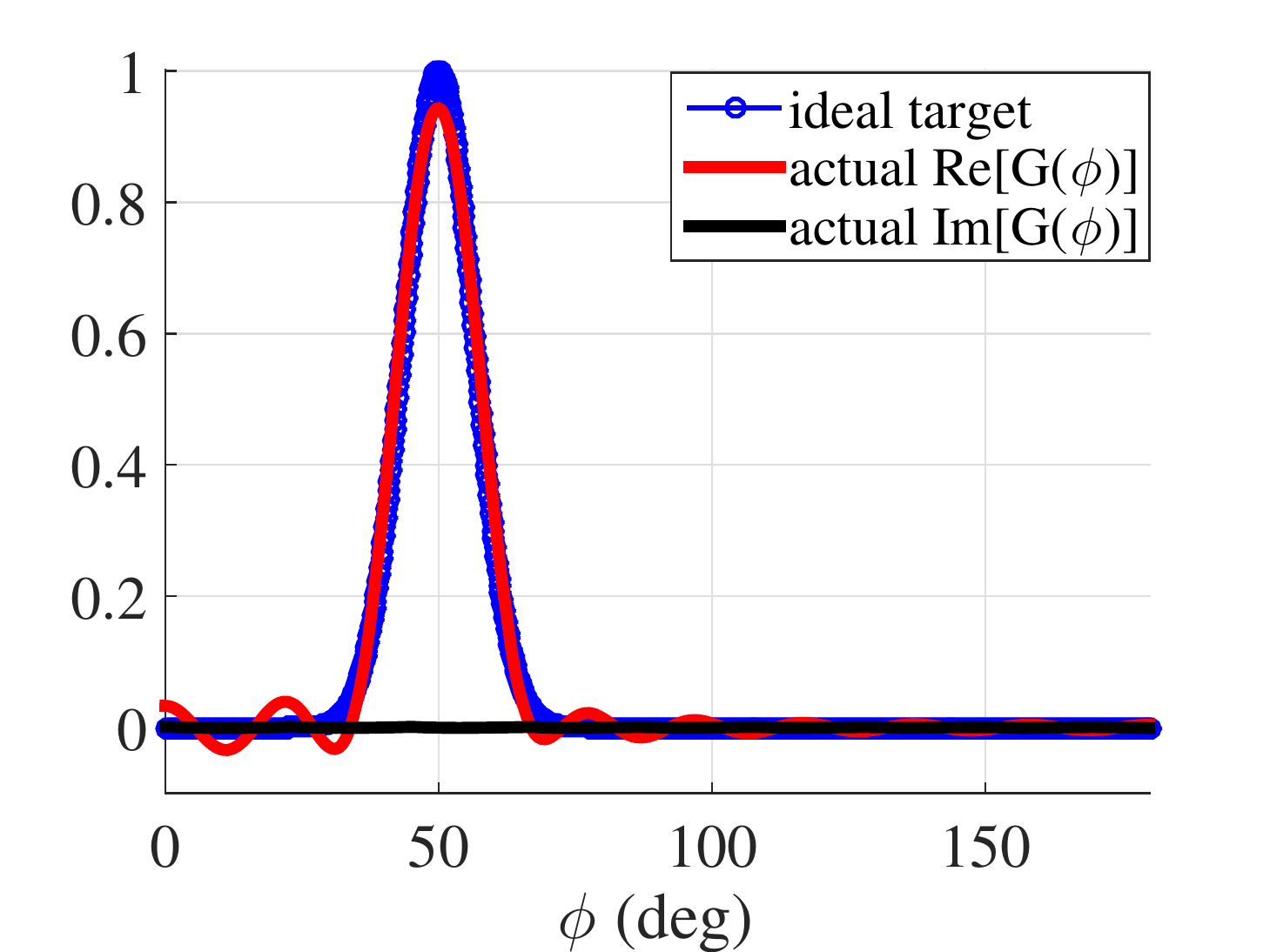}
   \label{fig:Fig8b}}
\subfigure[]{\includegraphics[width=4.2cm]{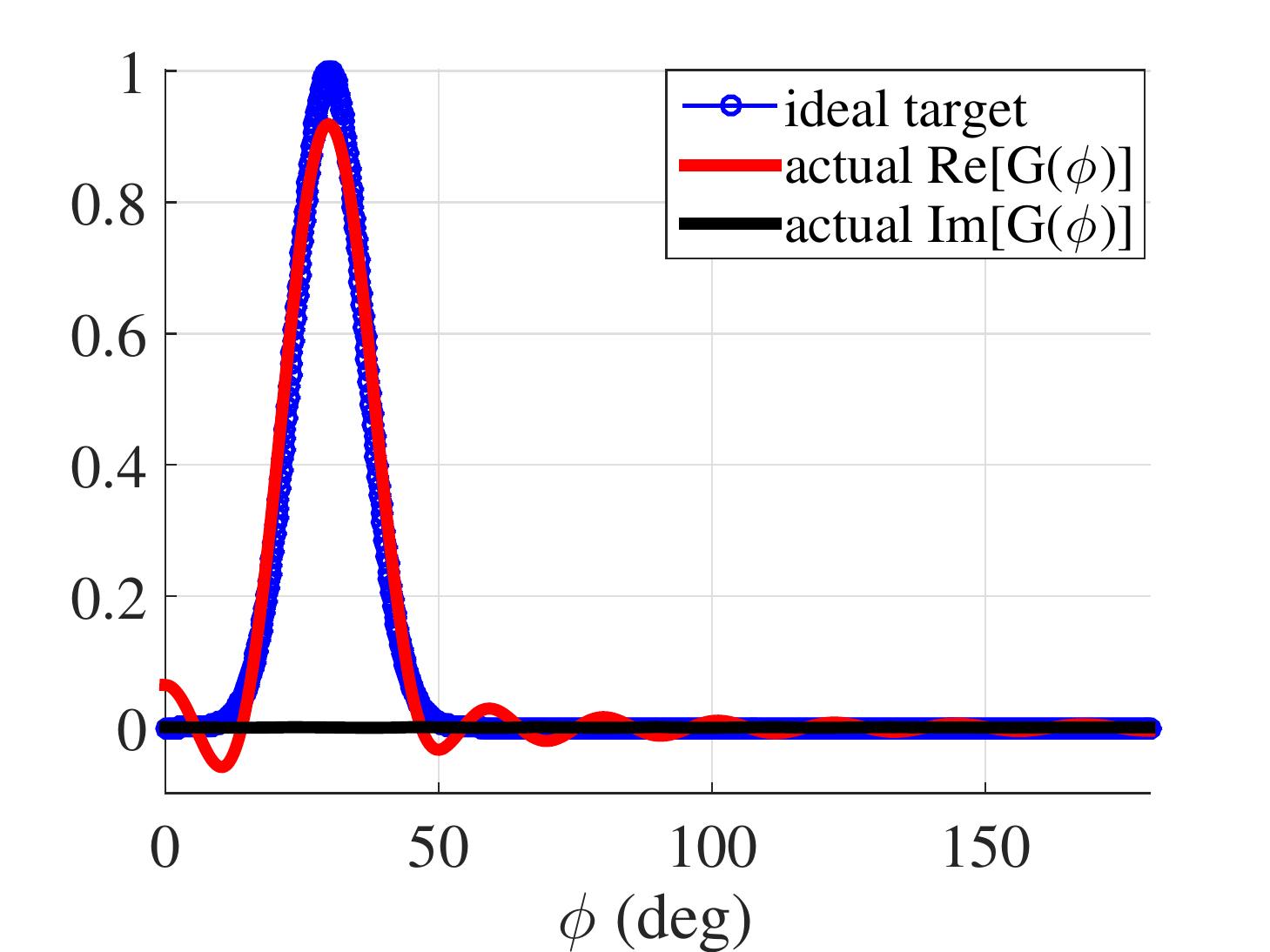}
   \label{fig:Fig8c}}
\subfigure[]{\includegraphics[width=4.2cm]{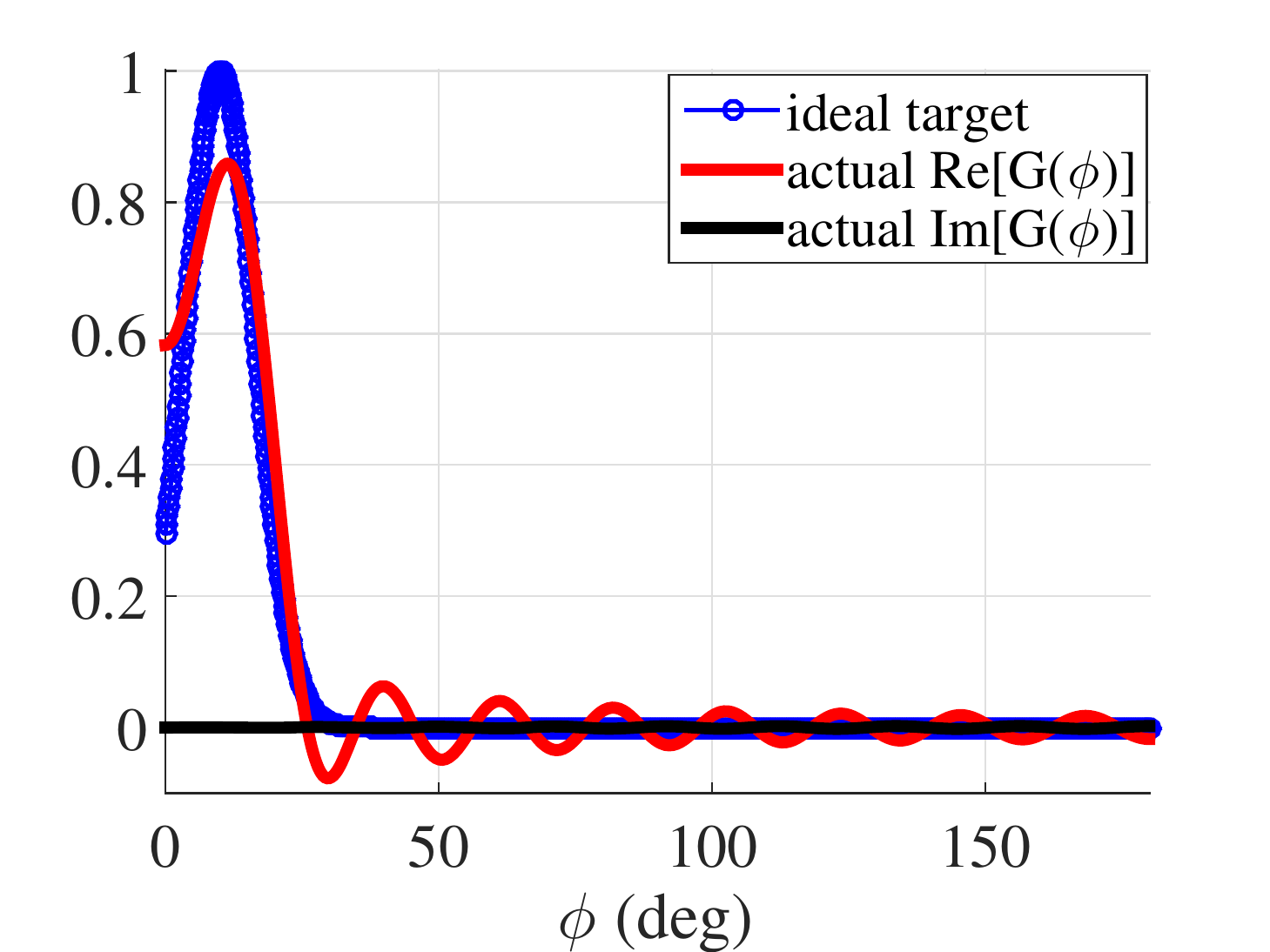}
   \label{fig:Fig8d}}
\caption{The ideal target $\tilde{G}(\phi)$ and the optimal actual pattern $G(\phi)$ (both real and imaginary parts) as functions of azimuthal angle $\phi$ for: (a) $\vartheta=80^{\circ}$, (b) $\vartheta=50^{\circ}$, (c) $\vartheta=30^{\circ}$, (d) $\vartheta=10^{\circ}$. Each optimal pattern corresponds to a different set of output fields dictated by the target pattern. Plot parameters: $\gamma=40$, referring to (\ref{eq:DesiredPatternFamily2}), $k_0L=0.1$ and $M=80$.}
\label{fig:Figs8}
\end{figure}

In Figs \ref{fig:Figs8}, we show how the emitter array used in Figs \ref{fig:Figs7} performs when the target pattern of (\ref{eq:DesiredPatternFamily2}) is quite directive $(\gamma=40)$ and tilted with various tilt angles $\theta$. The same coloring rules as in Figs \ref{fig:Figs6} apply. The only difference between the systems producing the responses of Figs \ref{fig:Figs8} is the fields $F_m$, $m=-N,\cdots, N$ which are obtained by inverting the aforementioned linear system $\textbf{S}\cdot \textbf{f}=\textbf{v}$ with different constant vectors $\textbf{v}$ computed from (\ref{eq:ConstantVector}). It is remarkable that the performance of our method deteriorates when the target beam gets less parallel to the interface $y=0$ (smaller angle $\theta$) despite the fact that $u_{max}$ is similar in all the four considered scenarios. To put it alternatively, the approximate inequality (\ref{eq:RuleOfThumb}) defining the range of lattice period $k_0L$ for successful implementation of the described technique, obviously does not take into account all the parameters of the considered problem. A tilted target radiation pattern $\tilde{G}(\phi)$ contains an additional inherent difficulty since it forces stronger output fields from the emitters close to the one end of the array. Such an imbalance in the distribution of the optimal fields $F_m$ with $m=-M,\cdots, M$ along the lasers at $y=0$ leads to a less accurate mimic of the target pattern; naturally, the highest performance is recorded for $\theta=90^{\circ}$ as in Fig. \ref{fig:Fig7c}. Note that the depicted results are not the best we can achieve with the proposed method; both in Figs \ref{fig:Figs7} and \ref{fig:Figs8}, we keep the optical distance between two consecutive lasers fixed (and quite small): $k_0L=0.1$. We did not calibrate this parameter in order to obtain perfect results as in Fig. \ref{fig:Fig6c} since our intention was to demonstrate the validity of (\ref{eq:RuleOfThumb}) and express clearly the limitations of the presented concept when more and more challenging targets are posed.

\section{Conclusions and Future Work}
\label{ConcFut}
In this work, we present an inverse method of picking (by solving a linear system) the optimal local fields at the outputs of a periodic laser array in order to make a desired far-field pattern. We concluded that the quality of the reconstruction of a given beam is directly related to the distance between two emitters. For this reason, we obtained an elegant inequality that suggests a safe range for that crucial parameter of our consideration. It is remarkable that for the successful implementation of the proposed technique, is not necessary to have a well-conditioned matrix; the numerical inversion is carried out without stability issues due to certain advantageous features of the matrix (it is real, symmetric and defined solely through Bessel function of zeroth order). 

In our analysis, we have assumed that creating a specific set of local fields at the ends of the laser waveguides is always feasible. However, such a result is related to the characteristics of the waveguides (length, cavity coupling), the employed materials (gain, non-linearities) and the current driving (injection, sources) \cite{NewChoquette}. A significant step forward would be to formulate the inverse problem not with respect to the outputs of the emitters (which is done in this work) but with respect to the geometrical, material and power features of the laser arrays which produce a far-field target response.

Another very interesting continuation of the present work would be to consider the problem of inversely developing a far-field pattern dependent not only on the azimuthal angle but also on the zenith angle of spherical coordinates. This would require a two-dimensional array of emitters and the electric field pattern would possess two components normal to the radial direction.

\section*{Acknowledgement}
This work has been partially supported by Ministry of Education and Science of the Republic of Kazakhstan via Contract number 339/76-2015 and Nazarbayev University ORAU grants. The authors would like to thank Dr. Arje Nachman for bringing to their attention the utility of level set methods.

\end{document}